\documentclass{ws-procs9x6}

\begin{document}

\title{Quantum excitation-free radiation emission including multiple scattering}

\author{ULRIK I. UGGERH{\O}J}

\address{Department of Physics and Astronomy, University of Aarhus \\ 
 DK-8000 Aarhus C, Denmark}  


\maketitle

\abstracts{
In order to increase the luminosity of electron-positron colliders it is desirable to find a means to reduce the phase-space of the beams. The transverse cooling of positrons imposed by the quantum excitation-free radiation emission in a single crystal is considered as a potential route to achieving ultra-cold beams. An analysis of the problem is presented, including an evaluation of the contribution from multiple scattering during the passage. The analysis shows that an emittance reduction may be achieved in special cases, but in general the emittance will increase as a result of the multiple scattering.
}

\section{Introduction}
In a series of theoretical papers, the quantum excitation-free radiation emission in a single crystal has been discussed \cite{Huan95,Huan96}. It is shown that the transverse action - and thereby the emittance - decreases exponentially towards the minimum value $\hbar/2$, corresponding to an emittance of half of the Compton wavelength. This applies as long as the radiation is in the undulator regime where the angle of emission is larger than the pitch angle. 
On the other hand, experiments \cite{Baur97} have been performed which in agreement with theoretical expectations based on a completely different analysis show that positrons in contrast to electrons generally suffer heating instead of cooling. The present work is motivated by two things: The desire to estimate the potential of single crystals as `devices' for the production of ultra-cold beams and a wish to find a consensus between two apparently different theoretical approaches and experimental results.
In the following the outline of \cite{Huan96} is followed, with the inclusion of multiple scattering, and it is shown that for the experiment one should indeed expect heating as concluded from both theories.
 Finally, it is shown that in special cases it is expected that transverse cooling can be achieved.

\section{Channeling, multiple scattering and dechanneling}
The large fields present near the nuclei in solid materials may in the case of
single crystals add coherently such that a penetrating particle experiences a
continuous field along its direction of motion - the so-called continuum approximation~\cite{Lind65}. If further the particle is
incident with a sufficiently small angle to a particular crystallographic
direction, inside the so-called Lindhard angle, the negatively/positively
charged particle is constrained to
move near/far from the nuclei and the electron clouds surrounding these. This
is the channeling phenomenon~\cite{Lind65} which has found widespread
applications in physics. For a general introduction to channeling and applications, see eg.\ \cite{Sore89}.
The critical angle for planar channeling is given by 
\begin{equation}
\psi_p=\sqrt{4Z_1Z_2e^2Nd_pCa_\mathrm{TF}/pv}
\end{equation}
where $Z_1e$ is the charge of the penetrating particle, $Z_2e$ that of the lattice nuclei, $p$ the momentum, $v$ the velocity, \(Nd_p\) is the planar density of atoms, \(N\) being the atomic density and \(d_p\) the planar spacing, \(C\simeq\sqrt{3}\) is Lindhard's constant and \(a_\mathrm{TF}\) is the Thomas-Fermi screening distance.
\begin{figure}[htb!]
\begin{center}
\mbox{\includegraphics[width=9cm]{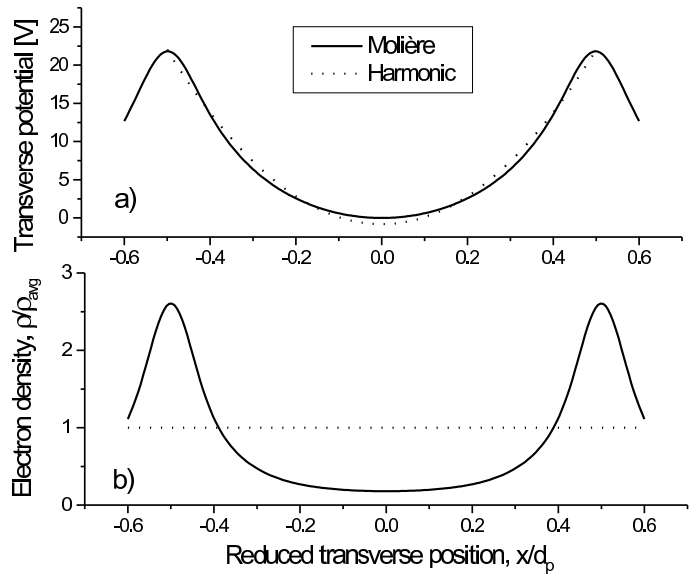}}
\caption[]{\textsl{The figure shows as a function of reduced transverse position, $x/d_p$, a) the transverse potential in the Moli$\grave{\mathrm{e}}$re and harmonic approximations and b) the electron density calculated from Poisson's equation.}
\label{fig:potential}}
\end{center}
\end{figure}

In the continuum approximation, the resulting transverse potential leads to a continuously focusing environment in which photon emission may take place without recoil to the emitting particle, the recoil being absorbed by the lattice. This is the so-called 'semi-classical channeling radiation reaction' \cite{Huan95,Huan96}.
In the previous papers on this phenomenon two main assumptions are made: The particle is moving in a harmonic potential and the energy of the photons emitted is small compared to the energy of the particle. Disregarding for the moment the potentially important case of axially channeled positrons along a crystal axis, this leaves only planar channeled positrons since channeled electrons are in a strongly anharmonic potential. Axially channeled positrons may be confined to a region between certain strings if their transverse energy is very low, so-called proper channeled positrons. In this case the transverse potential can be well approximated by a harmonic potential, see e.g.\ \cite{Lerv67}.
As seen from figure \ref{fig:potential}a, the harmonic approximation is clearly
well suited for positive particles.
\begin{figure}[htb!]
\begin{center}
\mbox{\includegraphics[width=9cm]{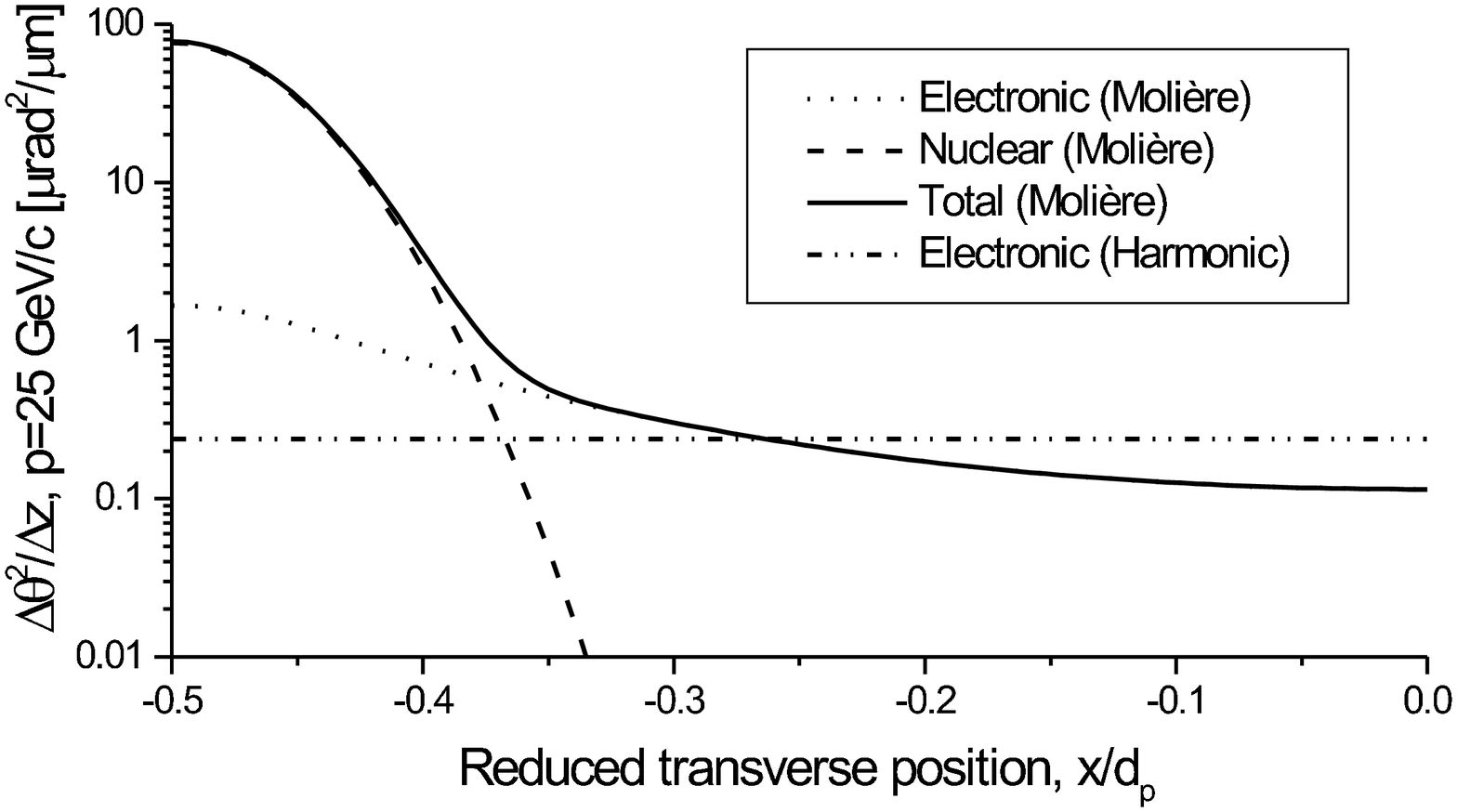}}
\caption[]{\textsl{The squared scattering angle per unit distance, $\Delta\theta^2/\Delta z$, as a function of reduced transverse position, $x/d_p$, for a $p=25$ GeV/c beam in silicon.}
\label{fig:mulscat}}
\end{center}
\end{figure}

The transfer of channeled particles to states above the barrier, so-called random, is referred to as dechanneling.
The length, \(L_\mathrm{D}\), over which a planar channeled beam of protons has been reduced to the fraction 1/e of the initial intensity by transfer to the random beam is given for \(\gamma\gg 1\) by~\cite{Biry94a}, \cite{Biry94b}:
\begin{equation}
L_\mathrm{D}=\frac{256}{9\pi^2}\frac{pv}{\ln(2\gamma mc^2/I)-1}\frac{a_\mathrm{TF}d_p}{Z_1e^2}
\label{dechlen}
\end{equation}
where \(I\) is the ionisation potential.
Eq.\ (\ref{dechlen}) has been shown to be in good agreement with measured values of \(L_\mathrm{D}\) at room temperature~\cite{Biry94b}. 
Due to the dependence of electron density on transverse position, see figure \ref{fig:potential}a, the dechanneling process which is a result of the multiple scattering depends itself on the transverse position and therefore on the transverse energy.
In figure \ref{fig:mulscat} is shown the squared scattering angle per unit distance, $\Delta\theta^2/\Delta z$, for a $p=25$ GeV/c beam in silicon. For low transverse energies, the multiple scattering is dominated by the interaction with electrons and it is observed that $\Delta\theta^2/\Delta z$ is almost a factor of two lower for the more accurate Moli$\grave{\mathrm{e}}$re approximation than for the harmonic potential (for the Moli$\grave{\mathrm{e}}$re approximation, see e.g.\ \cite{Biry97}).

\section{Quantum excitation-free radiation emission including multiple scattering}
Following \cite[eqs.\ (30) and (31)]{Huan96} the time evolution of the longitudinal Lorentz factor, $\gamma_z\equiv\sqrt{m^2c^4+p_z^2c^2}/mc^2=E_z/mc^2$, in a continuous focusing environment is given by the differential equation
\begin{equation}
\frac{\mathrm{d}\gamma_z(t)}{\mathrm{d}t}=-\Gamma_cG\gamma_z^{3/2}J_x(t)
\label{diffeq1}
\end{equation}
i.e.\ it couples to the transverse action,
 $J_x=E_x/\omega_z$, which evolves as
\begin{equation}
\frac{\mathrm{d}J_x(t)}{\mathrm{d}t}=-\Gamma_cJ_x-\frac{3}{4}\Gamma_cG\gamma_z^{1/2}(t)J_x^2(t)
\label{diffeq2}
\end{equation}
leading to two coupled differential equations, both in the absence of multiple scattering. Here $\Gamma_c=2r_eK/3mc$ is given by the focusing parameter, $K$, related to the transverse potential height, $U_0$, and planar spacing, $d_p$, as $U_0=K(d_p/2)^2/2$ and $G=\sqrt{K/m^3c^4}$ is a convenient constant expressing the focusing strength.

The solution is given as
\begin{equation}
J_x(t)=J_{x0}(1+\frac{5}{8}\gamma_{z0}^2\theta_{p0}^2(1-\exp(-\Gamma_ct)))^{-3/5}\exp(-\Gamma_ct)
\label{solution1}
\end{equation}
and
\begin{equation}
\gamma_z(t)=\gamma_{z0}(1+\frac{5}{8}\gamma_{z0}^2\theta_{p0}^2(1-\exp(-\Gamma_ct)))^{-4/5}
\label{solution2}
\end{equation}
where $\theta_{p0}=\sqrt{2GJ_{x0}/\gamma_{z0}^{3/2}}$ is the initial pitch angle, $J_{x0}$ the transverse action and $\gamma_{z0}$ the longitudinal Lorentz factor upon entry.

To include multiple scattering we use the analysis for the dechanneling length, $L_{\mathrm{D}}$, of positive particles with $\gamma\gg1$. The dechanneling process arises due to a steady increase of the transverse energy imposed by multiple scattering, i.e.\ the transverse energy increases as $\mathrm{d}E_x/\mathrm{d}z=U_0/L_{\mathrm{D}}$. Therefore the transverse action increases as $\mathrm{d}J_x/\mathrm{d}t=U_0c/L_{\mathrm{D}}\omega_z$, but since the effective dechanneling length depends on the transverse energy approximately as $L_{\mathrm{D}}=L_{\mathrm{D0}}U_0/J_x\omega_z$, the transverse action changes according to
\begin{equation}
\mathrm{d}J_x/\mathrm{d}t=c/L_{\mathrm{D0}}J_x
\label{transact}
\end{equation}
Here $L_{\mathrm{D0}}$ denotes the dechanneling length for states where $E_x\simeq U_0$. This is found by dividing the squared scattering angle for $x=0$ by the average over $x$ of the squared scattering angles shown in figure and multiplying the dechanneling length from eq. (\ref{dechlen}) by this ratio, i.e.
\begin{equation}
L_{\mathrm{D0}}=L_{\mathrm{D}}d_p\Delta\theta^2(d_p/2)/2\int_0^{d_p/2}\Delta\theta^2(x)\mathrm{d}x
\label{ld0}
\end{equation} 
Thus, by combining eqs.\ (\ref{diffeq2}) and (\ref{transact}), the result is
\begin{equation}
\frac{\mathrm{d}J_x(t)}{\mathrm{d}t}=(\frac{c}{L_{\mathrm{D0}}}-\Gamma_c)J_x-\frac{3}{4}\Gamma_cG\gamma_z^{1/2}(t)J_x^2(t)
\end{equation}
whereas $\gamma_z(t)$ remains unaffected. By a change of variables $\Gamma_c'=\Gamma_c(1-c/\Gamma_cL_{\mathrm{D0}}$ and $G'=G(1-c/\Gamma_cL_{\mathrm{D0}})^{-1}$ the same type of coupled differential equations as eqs.\ (\ref{diffeq1}) and (\ref{diffeq2}) are obtained with solutions given by eqs.\ (\ref{solution1}) and (\ref{solution2}) with $\Gamma_c'$ and $G'$ instead of $\Gamma_c$ and $G$ and with a new pitch angle, $\theta_{p0}'=(1-c/\Gamma_cL_{\mathrm{D0}})^{-1/2}\theta_{p0}$. To get the true pitch angle as a function of time the inverse transformation is applied, i.e.\ $\theta_{p}=(1-c/\Gamma_cL_{\mathrm{D0}})^{1/2}\theta_{p}'$ where $\theta_{p}'=\sqrt{2G'J_{x}'/\gamma_{z}'^{3/2}}$ is the modified pitch angle. As expected and seen from the results below, a good measure of the depth at which the cooling effect starts to appear is given by $c\tau_c=c/\Gamma_c$.

\section{Results}

\begin{figure}[htb!]
\begin{center}
\mbox{\includegraphics[width=9cm]{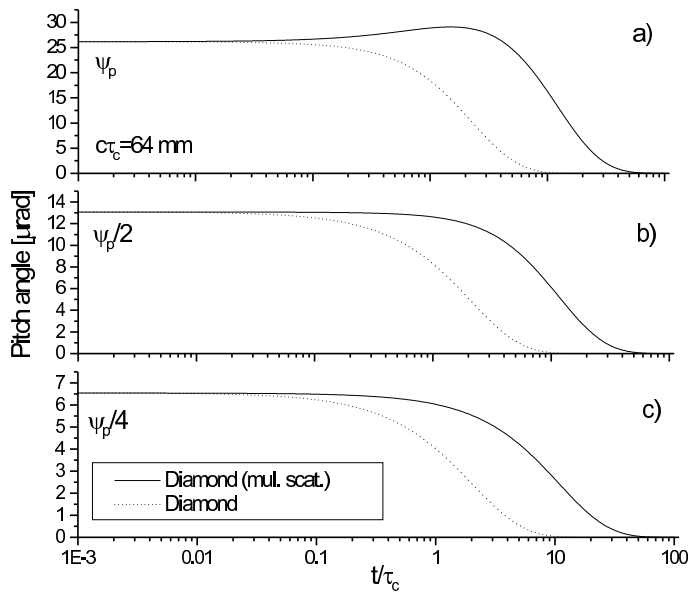}}
\caption[]{\textsl{The pitch angle in diamond (110) as a function of normalized penetration time, $t/\tau_c$. The full drawn curve is calculated including multiple scattering while the dotted line excludes this contribution. The graphs a), b) and c) are for incidence angles $\psi_p$, $\psi_p/2$ and $\psi_p/4$, respectively.}
\label{fig:diamond}}
\end{center}
\end{figure}

In figure \ref{fig:diamond} is shown the pitch angle calculated as a function of normalized penetration time, $t/\tau_c$, for 25 GeV positrons in a (110) diamond, with and without inclusion of the multiple scattering. It is seen that while the cooling starts around $t/\tau_c=0.1$ without multiple scattering, it is postponed to values above $t/\tau_c\simeq5$ when this additional effect is taken into account. Since there is always an incoherent contribution to the radiation emission which typically takes place as in an amorphous medium, it is important to note the scale of $c\tau_c$ compared to the amorphous radiation length, $X_0$, which for diamond is 122 mm. It is thus not possible to utilize planar channeling in diamond for cooling of a 25 GeV beam of positrons for an angle of incidence near $\psi_p$. However, for smaller angles the cooling starts already at $\simeq0.1t/\tau_c$, i.e.\ a 10 mm thick diamond would suffice to initiate the cooling.

\begin{figure}[htb!]
\begin{center}
\mbox{\includegraphics[width=9cm]{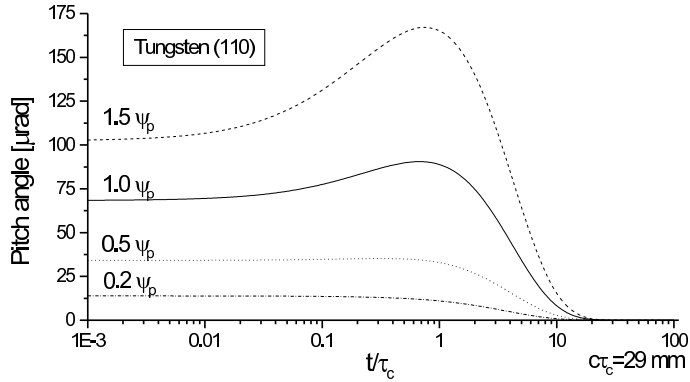}}
\caption[]{\textsl{The pitch angles in tungsten (110) as a function of normalized penetration time, $t/\tau_c$, for 4 different angles of incidence as indicated. The positron energy is set to 25 GeV and all curves are calculated including multiple scattering.}
\label{fig:tungsten_angle}}
\end{center}
\end{figure}

In figure \ref{fig:tungsten_angle} is shown the different behaviours for several angles of incidence, calculated for tungsten (110). Strong cooling is found to appear early for small values of the angle of incidence, but for tungsten $c\tau_c\simeq8X_0$ which means that even for small angles there will be a strong influence from incoherent scattering.

\begin{figure}[htb!]
\begin{center}
\mbox{\includegraphics[width=9cm]{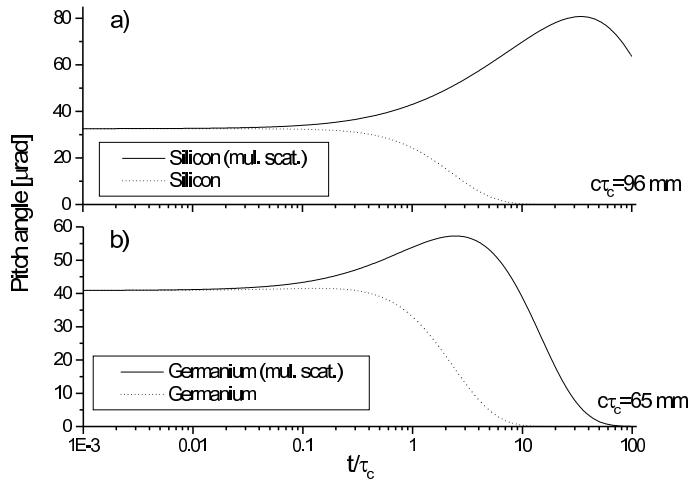}}
\caption[]{\textsl{The pitch angle in a) silicon and b) germanium (110) as a function of normalized penetration time, $t/\tau_c$. In each case, the initial pitch angle equals the planar critical angle, $\psi_p$, and the full curves are calculated including multiple scattering while the dotted curves do not include this effect.}
\label{fig:si_ge}}
\end{center}
\end{figure}

In figure \ref{fig:si_ge} is shown results for silicon and germanium showing that for light materials, where the lattice is not very compact as in diamond, the influence of multiple scattering is much stronger than for heavier ones.

\begin{figure}[htb!]
\begin{center}
\mbox{\includegraphics[width=9cm]{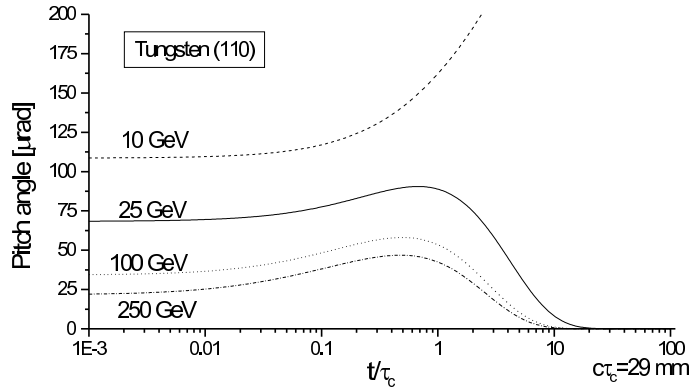}}
\caption[]{\textsl{The pitch angle in tungsten (110) as a function of normalized penetration time, $t/\tau_c$, for 4 different energies as indicated. In each case, the initial pitch angle equals the planar critical angle, $\psi_p$, and all curves are calculated including multiple scattering.}
\label{fig:tungsten_energy}}
\end{center}
\end{figure}

To get an impression of the variation of the cooling effect as a function of energy in figure \ref{fig:tungsten_energy} is shown the results for 4 different energies. Clearly, as the energy increases, the influence of multiple scattering diminishes and cooling starts earlier.

\begin{figure}[htb!]
\begin{center}
\mbox{\includegraphics[width=9cm]{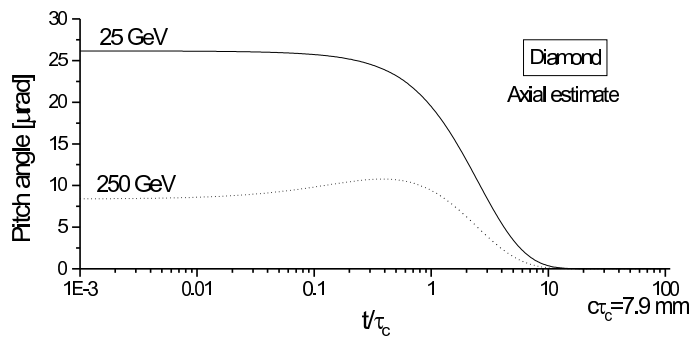}}
\caption[]{\textsl{The estimated axial pitch angle in diamond as a function of normalized penetration time, $t/\tau_c$, for 2 different energies as indicated. In each case, the initial pitch angle equals the planar critical angle, $\psi_p$, and both curves are calculated including multiple scattering.}
\label{fig:diamond_axial}}
\end{center}
\end{figure}

As a crude estimate, the axial potential can be approximated by a harmonic potential with barrier height equal to that found by the Moli$\grave{\mathrm{e}}$re potential. The two kinds of motions in the planar and axial channeling cases are very different and it is not expected to get anything but an indication of the magnitude of the effect from this analysis. However, in figure \ref{fig:diamond_axial} is shown the results obtained by this procedure for two energies and we note that in this case the characteristic cooling length is \emph{much} shorter than $X_0$. This indicates that the much stronger axial fields may indeed provide substantial cooling.

\begin{figure}[htb!]
\begin{center}
\mbox{\includegraphics[width=9cm]{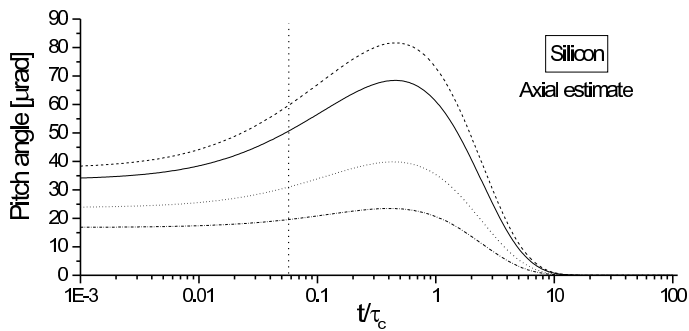}}
\caption[]{\textsl{The estimated axial pitch angle in silicon as a function of normalized penetration time, $t/\tau_c$, for 4 different angles of incidence. All curves are calculated including multiple scattering.}
\label{fig:cool_na43}}
\end{center}
\end{figure}

Such an experiment has been performed by NA43 at CERN, see \cite{Baur97}, and compared to a theoretical analysis by Kononets, see \cite{Kono98}. It was found, in agreement with this theory, that positrons suffered heating in contrast to electrons.
In figure \ref{fig:cool_na43} is shown results under the same crude estimation method for the axial case as mentioned above. The actual value of the crystal thickness in the experiment is given by the vertical dashed line and it is seen - as observed in the experiment - that for all angles of incidence positrons suffer heating.

In \cite{Baie97} calculations of the transverse cooling including multiple scattering have also been performed. However, in order to present a fully self-consistent model, the authors have chosen to model the electron density by solving the Poisson equation for the harmonic potential. This yields an electron density which is constant as a function of transverse position and may underestimate the net cooling effect as a cause of this. On the other hand, in \cite{Baie97} scattering `on fluctuations of the planar potential', i.e.\ the nuclear contribution is not taken into account either. 
The analysis presented here supplements that of Baier and Katkov \cite{Baie97} in producing essentially the same conclusion by use of a different approach: Only under special circumstances may a penetrating particle experience a net cooling effect.

\section{Conclusions}
Qualitative agreement among two different theoretical approaches and an experiment on radiative angular cooling is shown. This gives considerable confidence in the predictions, especially of the semi-classical channeling radiation reaction approach. Furthermore, for this approach multiple scattering is included to obtain a more realistic estimate of the cooling properties. Rough estimates show that for existing positron beams a strong cooling effect may be achievable by means of axial channeling in a $\simeq20$ mm thick diamond crystal.
There are however still open questions which deserve attention, among others that of dilution of the longitudinal phase-space due to the severe straggling in energy loss and the possibility of cooling simultaneously in both transverse directions by proper axial channeling or channeling in carbon nanotubes \cite{Taji03}.

\section{Acknowledgments}
I am grateful for having benefitted substantially from encouraging discussions with P. Chen, T. Tajima, H. Rosu and J. Ellison. Furthermore, I wish to thank V. Baier for drawing my attention to the paper \cite{Baie97}. A 'Steno' grant from the Danish Natural Science Research Council is  gratefully acknowledged.

 Finally, I wish to thank Pisin Chen for his kind invitation to attend a workshop with a (for me) hitherto unseen fraction of visionary physicists.


\begin{thebibliography}{99}

\bibitem{Huan95}
	Z. Huang, P. Chen and R.D. Ruth, Phys. Rev. Lett. \textbf{74}, 1759 (1995)

\bibitem{Huan96}
	Z. Huang, P. Chen and R.D. Ruth, Nucl. Instr. Meth. B \textbf{119}, 192 (1996)

\bibitem{Baur97}
        A. Baurichter \textit{et al.}, Phys. Rev. Lett. \textbf{79}, 3415 (1997)

\bibitem{Lind65}
        J. Lindhard, Mat. Fys. Medd. Dan. Vid. Selsk. \textbf{34}, 1 (1965)

\bibitem{Sore89}
        A.H. S{\o}rensen and E. Uggerh{\o}j, Nucl. Sci. Appl. \textbf{3}, 147 (1989)

\bibitem{Lerv67}
        Ph. Lervig, J. Lindhard and V. Nielsen, Nucl. Phys. \textbf{A96}, 481 (1967)

\bibitem{Biry94a}
        V.M. Biryukov \textit{et al.}, Physics-Uspekhi \textbf{37}, 937 (1994)

\bibitem{Biry94b}
        V.M. Biryukov \textit{et al.}, Nucl. Instr. Meth. B \textbf{86}, 245 (1994)

\bibitem{Biry97}
        V.M. Biryukov, Y.A. Chesnokov and V.I. Kotov, Crystal Channeling and Its Application at High-Energy Accelerators, Springer-Verlag (1997)

\bibitem{Kono98}
 	Yu.V. Kononets, Nucl. Instr. Meth. B \textbf{135}, 40 (1998)       

\bibitem{Baie97}
	V.N. Baier and V.M. Katkov, Phys. Lett. A \textbf{232}, 456 (1997)

\bibitem{Taji03}
	T. Tajima, private communication, 2003

\end{thebibliography}
\end{document}